\documentclass[epj]{svjour} 

\usepackage{color}
\usepackage{graphics}
\usepackage{amsmath}
\usepackage{amssymb}
\begin{document}
\title{Transport of secondary electrons and reactive species in ion tracks}
\author{Eugene Surdutovich\inst{1,2}
    \and Andrey V. Solov'yov\inst{2,3}
}                     
\offprints{Eugene Surdutovich}          
\institute{Department of Physics, Oakland University, Rochester, Michigan 48309, USA
     \and
MBN Research Center,
60438 Frankfurt am Main, Germany
\and On leave from A.F. Ioffe Physical Technical Institute, 
194021 St. Petersburg, Russian Federation}
\date{Received: date / Revised version: date}
%
\abstract{ The transport of reactive species brought about by ions traversing tissue-like medium is analysed analytically. Secondary electrons ejected by ions are capable of ionizing other molecules; the transport of these generations of electrons is studied using the random walk approximation until these electrons remain ballistic. Then, the distribution of solvated electrons produced as a result of interaction of low-energy electrons with water molecules is obtained. The radial distribution of energy loss by ions and secondary electrons to the medium yields the initial radial dose distribution, which can be used as initial conditions for the predicted shock waves. The formation, diffusion, and chemical evolution of hydroxyl radicals in liquid water are studied as well.
} 
\maketitle

\section{Introduction}
\label{intro}

It is widely accepted that secondary particles such as electrons, solvated electrons, and free radicals play the key role in the radiation damage with ions~\cite{MSAColl,hyd2,Nikjoo97}. These particles are produced as a result of ionization and excitation of medium by ions as well as inelastic interactions of secondary electrons of the first and following generations. The entire scenario of radiation damage with ions, considered in the multiscale approach~\cite{MSAColl,pre}, includes the chemical stage when the above agents, formed in the vicinity of the ion's path, interact with nearby biomolecules. Then follows the predicted thermomechanical stage, during which the shock wave, generated by the gradient of energy deposited in the medium, may rupture the covalent bonds of biomolecules, and transport the reactive species formed near the ion's path to a much wider region. Then these species may interact with more distant biomolecules~\cite{MSAColl}.

The analysis of transport of secondary particles is important for all stages of this scenario. While the molecular dynamic simulation is the most adequate method of study for the latter stages~\cite{natnuke}, an analytical random walk method can be applied for the first stage~\cite{MSAColl}. In this paper, we elaborate the application of this method to describe the transport of two generations of electrons and reactive species such as free radicals and solvated electrons. We also analyse the location of energy deposition by secondary electrons and reconstruct the radial dose distribution based on the diffusion scenario.


\section{Transport of secondary particles originating on the ion's path}
\label{sec.tranaxis}

A large number of secondary electrons and radicals is produced in the immediate vicinity of the ion's path. The secondary electrons are ejected as a result of ionizations of the molecules of the medium by projectiles. The radicals originate as a result of either molecule excitation by ions or as a result of hydrolysis due to local heating as a result of ion's passage. The  production of secondary particles on the ion's path is described by the number of produced particles of hue ``$j$'' per unit length of the ion's path, $dN_{pj}/d\zeta$. An additional subscript ``$p$'' indicates that these particles originate from the ion's path.

The tree-dimensional diffusion of secondary particles such as electrons and free radicals is described by the following equation,
\begin{eqnarray}
\frac{\partial n_{pj}({\bf r}, t)}{\partial t} = D_j \nabla^2 n_{pj}({\bf r}, t) - \frac{n_{pj}({\bf r}, t)}{\tau_j}, \label{raddiff}
\end{eqnarray}
where $n_{pj}({\bf r}, t)$ is the average number density of corresponding secondary particles at a location ${\bf r}$ (this vector connects points of origin of particles on the path and their observation), $D_j$ is the corresponding diffusion coefficient in the medium, $\tau_j$ is the average lifetime. For electrons, $D_j=D_e=vl/6$, where $v$ is the electron's velocity and $l$ is their elastic mean free path in the medium.

Equation~(\ref{raddiff}) has a well-known solution~\cite{MSAColl,Chandra}
\begin{eqnarray}
\frac {d n_{pj}(t, r)}{d\zeta}=\frac{dN_{pj}}{d\zeta} \left(\frac{1}{4\pi D_j t}\right)^{3/2}\exp\left(-\frac{r^2}{4 D_j t}-\frac{t}{\tau_j}\right),
\label{Prwalk}
\end{eqnarray}
where $r$ is the length of ${\bf r}$. Equation (\ref{Prwalk}) can be integrated over $\zeta$ to obtain the number density for each species originating at the ion's path,
\begin{eqnarray}
n_{pj}(t, \rho)=\frac{dN_{pj}}{d\zeta} \frac{1}{4\pi D_j t}\exp\left(-\frac{\rho^2}{4 D_j t}-\frac{t}{\tau_j}\right).
\label{PrwalkInt}
\end{eqnarray}

The flux of secondary particles, $d \Phi_{pj}$, through an element of a surface $d{\bf A}$ is given by
\begin{eqnarray}
d\Phi_{pj}=\int_{-\infty}^\infty \left(d{\bf A}\cdot \nabla D_j\frac {d n_{pj}(t, r)}{d\zeta}\right) d\zeta , \label{flux1}
\end{eqnarray}
where the integration is done only over $\zeta$, but not $d{\bf A}$. In order to calculate the flux of secondary particles through a cylindrical shell of radius $\rho$ and length $\xi$, with the ion's path as the axis, $d{\bf A}$ is chosen to be $-{\bf \rho} d\phi \xi$, where $\phi$ is the azimuthal angle. The gradient of {\em spherically} symmetric $\frac {d n_{pj}(t, r)}{d\zeta}$ is collinear with ${\bf r}$. Then, substituting the dot product ${\bf r}\cdot{\bf \rho}=r \rho \rho/r=\rho^2$ in (\ref{flux1}), we obtain
\begin{eqnarray}
d\Phi_{pj}=-\int_{-\infty}^\infty \left(d\phi \xi \rho^2 \frac{1}{r}\frac{\partial}{\partial r} D_j\frac {d n_{pj}(t, r)}{d\zeta}\right) d\zeta . \label{flux2}
\end{eqnarray}
Since this expression is cylindrically symmetric, there is no dependence of $\phi$, and $A$ can be taken as the area of a cylindrical shell of radius $\rho$ and length $\xi$. Then $\Phi_{pj}$ is the total flux of radicals through this shell.
\begin{eqnarray}
\Phi_{pj}=- 2\pi \rho^2 \xi D_j \int  \frac{1}{r}\frac{\partial}{\partial r}\left(\frac {d n_{pj}(t, r)}{d\zeta}\right)d\zeta  =\frac{2\pi \rho^2 \xi}{2t} \nonumber\\
   \times \int  \frac{dN_{pj}}{d\zeta} \left(\frac{1}{4\pi D_j t}\right)^{3/2}\exp\left(-\frac{r^2}{4 D_j t}-\frac{t}{\tau_j}\right)d\zeta
 .\label{flux3}
\end{eqnarray}
Since the characteristic spatial scale in radial direction is in nanometers and in the axial direction is micrometers, $\frac{dN_{pj}}{d\zeta}$ can be assumed to be constant. Then (\ref{flux3}) can be integrated to give the flux,
\begin{eqnarray}
\Phi_{pj}=\frac{\partial N_{pj}(\rho, t)}{\partial t}
   = \frac{dN_{pj}}{d\zeta} \frac{\rho^2 \xi}{4 D_j t^2}\exp\left(-\frac{\rho^2}{4 D_j t}-\frac{t}{\tau_j}\right), \label{flux4}
\end{eqnarray}
which can be further integrated over time to obtain the total number of secondary particles of hue $j$, produced on the path, incident on the cylindrical shell. Such a calculation (in this work, only shown for this example) is important for the finding of the probability of a lesion in biomolecules~\cite{MSAColl}.

\section{Transport of particles produced as a result of interactions of secondary particles with the medium}
\label{sec.tranontheway}

Many of secondary electrons ejected from the location of the ion's path have the energy sufficient for ionization of molecules of the medium. These ionizations take place at different distances from the path. Then, the ionized molecules are likely to become sources of free radicals. The latter also originate away from the path. After these ionizations, both the newly ejected electrons and those that ionize the molecules have lower energies and, therefore, shorter elastic mean free paths. Finally, radicals propagating from the path can react with the medium and produce new reactive species. All these effects can be described using the same approach, described in this section.



Each of these processes can be described by the coupled transport equations,
\begin{eqnarray}
\frac{\partial n_{pj}({\bf r}, t)}{\partial t} = D_{j} \nabla^2 n_{pj}({\bf r}, t) - \sum_i\frac{ n_{pj}({\bf r}, t)}{\tau_{ji}}~, \nonumber\\
\frac{\partial n_i({\bf r}, t)}{\partial t} = D_i \nabla^2 n_i({\bf r}, t) + \frac{n_{pj}({\bf r}, t)}{\tau_{ji}}-\frac{n_i({\bf r}, t)}{\tau_i}~,
\label{coupled}
\end{eqnarray}
where index ``$i$'' marks the a specific kind of newly formed reactive species. The first of equations (\ref{coupled}) describes the random walk of secondary particles ejected by the ion. This equation can be solved the same way as (\ref{raddiff}) to give
\begin{eqnarray}
n_e(t, \rho)=\frac{dN_e}{d\zeta}\frac{1}{4\pi D_e t}\exp\left(-\frac{\rho^2}{4 D_e t}-\frac{t}{\tau} \right),
\label{nesolint}
\end{eqnarray}
where $1/\tau=\sum 1/\tau_i$.

Then, the rest of equations (\ref{coupled}) become diffusion equations with sources, and they can be solved analytically as well with a use of the Green's functions, which are given by,
\begin{eqnarray}
G_i(t-t',{\bf r}-{\bf r'})=\left(\frac{1}{4\pi D_i (t-t')}\right)^{3/2}\nonumber \\ \times \exp\left(-\frac{({\bf r}-{\bf r'})^2}{4 D_i (t-t')}-\frac{t-t'}{\tau_i} \right).
\label{green}
\end{eqnarray}
The solutions are:
\begin{eqnarray}
n_i(t,{\bf r})=\int G(t-t',{\bf r}-{\bf r'})\frac{n_e(t',{\bf r'})}{\tau_{ij}} dt' d{\bf r'}.
\label{greensol}
\end{eqnarray}
Each of these solutions requires four integrations, which are rather bulky, but doable. In order to simplify these integrations, it is advisable to use cylindrical coordinates with $\zeta$-axis along the ion's path. The expression looks as,
\begin{eqnarray}
n_i(t,{\bf r})=\frac{1}{\tau_{ij}} \frac{dN_e}{d\zeta}\int \left(\frac{1}{4\pi D_i (t-t')}\right)^{3/2}\nonumber \\ \times \exp\left(-\frac{({\bf r}-{\bf r'})^2}{4 D_i (t-t')}-\frac{t-t'}{\tau_i} \right)\nonumber\\ \times \frac{1}{4\pi D_e t'}\exp\left(-\frac{\rho'^2}{4 D_e t'}-\frac{ t'}{\tau_{ij}} \right)dt' d{\bf r'},
\label{integrand0}
\end{eqnarray}
where
\begin{eqnarray}
({\bf r}-{\bf r'})^2=\rho^2-2\rho\rho'\cos(\phi-\phi')+\rho'^2+(\zeta-\zeta')^2.
\label{rmrp}
\end{eqnarray}
and $d{\bf r'}=\rho'd\rho'd\phi'd\zeta'$. Since $\zeta'$ dependence only appears in the exponential of (\ref{integrand0}) after substituting (\ref{rmrp}), this integration in infinite limits can be done first; and it is equal to $\sqrt{4 \pi D_i (t-t')}$. Similarly, the dependence on the azimuthal angle also only appears in the exponential of (\ref{integrand0}) after substituting (\ref{rmrp}), and
\begin{eqnarray}
\int \exp\left(\frac{2\rho\rho'\cos(\phi-\phi')}{4 D_i (t-t')}\right)d\phi'=2\pi I_0\left(\frac{\rho\rho'}{2 D_i (t-t')}\right),
\label{azim}
\end{eqnarray}
where $I_0$ is a Bessel function. After this, two more integrals are remaining,
\begin{eqnarray}
n_i(t,{\bf r})=\frac{1}{\tau_{ij}}\frac{dN_e}{d\zeta}\int \frac{1}{8\pi D_i D_e (t-t') t'}\nonumber \\ \times\exp\left(-\frac{\rho^2+\rho'^2}{4 D_i (t-t')}-\frac{t-t'}{\tau_i} -\frac{\rho'^2}{4 D_e t'}-\frac{ t'}{\tau_{ij}}  \right)\nonumber \\
I_0\left(\frac{\rho\rho'}{2 D_i (t-t')}\right)dt' \rho'd\rho'.
\label{integrand1}
\end{eqnarray}
First, the integral over $\rho'$ is of the kind
\begin{eqnarray}
\int_0^\infty\exp\left(-\frac{\rho'^2}{a^2}\right) I_0\left(\frac{\rho'}{b}\right) \rho'd\rho'=\frac{a^2}{2}\exp\left(\frac{a^2}{4 b^2}\right).
\label{integrand2}
\end{eqnarray}
After substituting the expressions for $a$ and $b$ into (\ref{integrand2}), (\ref{integrand1}) becomes
\begin{eqnarray}
n_i(t,\rho)=\frac{1}{4\pi\tau_{ij}} \frac{dN_e}{d\zeta}\int_0^t \frac{1}{D_e t'+D_i(t-t')}\nonumber \\ \times \exp\left(-\frac{\rho^2}{4( D_e t'+D_i(t-t'))}-\frac{t-t'}{\tau_i}- \frac{t'}{\tau_{ij}}\right)dt'.
\label{integrand3}
\end{eqnarray}
The last integration is not analytic, but it can be readily done numerically. Thus, $n_i$ is obtained as a function of time and the distance from the axis; it can be added to $n_{pi}$ for a given species, given by   (\ref{PrwalkInt}).

The contribution to the flux due to $\nabla n_i$ can also be calculated as a function of $\rho$ and $t$. All these calculations include parameters $dN_{pi}/d\zeta$, $\tau_i$, and $\tau_{ij}$ for each species of interest.


\section{Transport of electrons ejected by secondary electrons}
\label{sec.tertiaryelectrons}

Many of the secondary electrons ejected have enough energy to ionize water molecules in the medium. Indeed, more than 65\% of these electrons have energies higher than the ionization potential of water molecules. The second ``wave'' of ionization has been discussed~\cite{epjd}, but only from the point of view of remaining energy, and it was concluded that the third wave is insignificant. For some applications, it is sufficient to consider the diffusion of secondary electrons leaving the second wave aside~\cite{epjdisacc2011}. However, Sec.~\ref{sec.tranontheway} suggests an analytic method for accounting of the second generation of electrons.

\begin{figure}
\begin{centering}
\resizebox{0.99\columnwidth}{!}
{\includegraphics{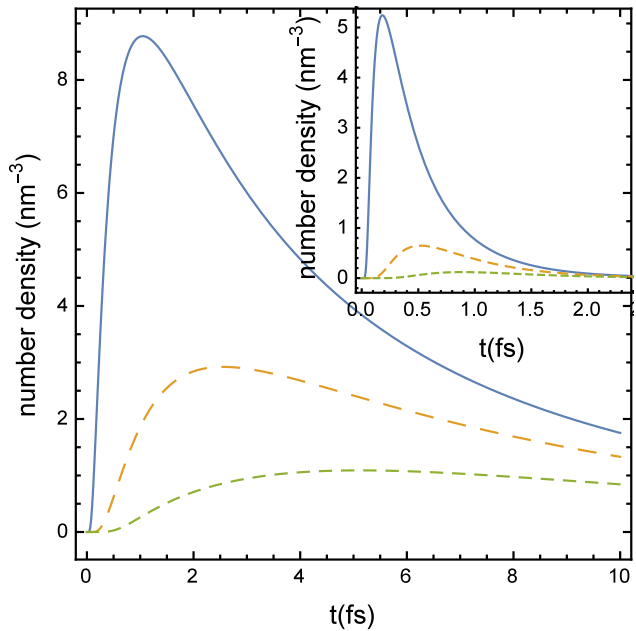}}
\caption{\label{fig:ele} (Colour online)
Number densities of secondary electrons of second generation at 0.5- (solid) 1-, and 1.5-nm distances (dashed with decreasing dash size, respectively) from the path as functions of time. The distributions of the first generation at the same distances are shown in the inset.}
\end{centering}
\end{figure}
\begin{figure}
\begin{centering}
\resizebox{0.99\columnwidth}{!}
{\includegraphics{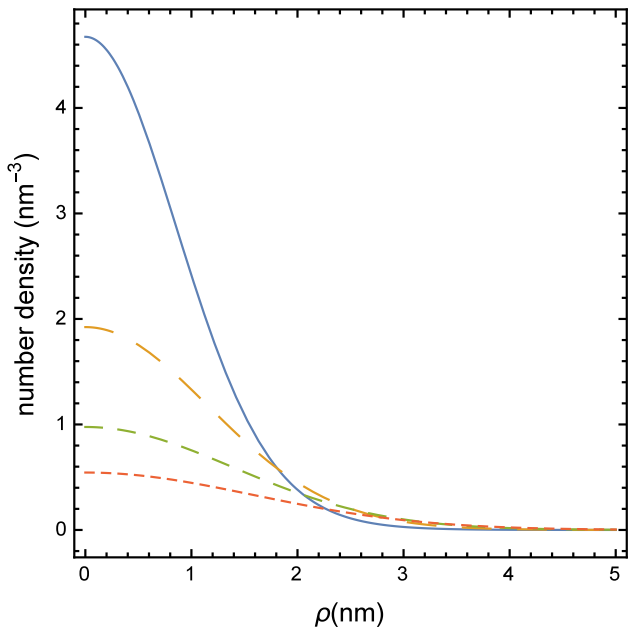}}
\caption{\label{fig.el2} (Colour online)
Number densities of secondary electrons after they have ionized water molecules at 5, 10,  15, and 20-fs times (solid and dashed lines with a diminishing dash size, correspondingly) as functions of the distance from the ion's path. }
\end{centering}
\end{figure}

The formation and transport of the second wave ionization can be described by equations~(\ref{coupled}),
\begin{eqnarray}
\frac{\partial n_1({\bf r}, t)}{\partial t} = D_1 \nabla^2 n_1({\bf r}, t) - \frac{n_1({\bf r}, t)}{\tau_1}, \nonumber\\
\frac{\partial n_2({\bf r}, t)}{\partial t} = D_2 \nabla^2 n_2({\bf r}, t)\nonumber\\ + \frac{n_1({\bf r}, t)}{\tau_1}-\frac{n_{2}({\bf r}, t)}{\tau_{2}}.
\label{coupledE}
\end{eqnarray}
Here, index ``$1$'' marks secondary electrons of the first generation. Their energy is taken to be equal to be the average energy of secondary electrons formed in the vicinity of the Bragg peak, i.e., 45~eV. Hence, $D_1=vl/6=0.265$nm$^2$fs$^{-1}$ and $\tau_1=l_{ion}/v=0.64$fs (all mean free path data are taken from Refs.~\cite{Nikjoo1998,Tung}. Index ``$2$'' corresponds to 15-eV electrons, formed as a result of the second wave of ionization by secondary electrons; $D_2=0.057$nm$^2$fs$^{-1}$ and $\tau_2=15.3$fs. The solutions of the above equations are obtained similarly to those of ~(\ref{coupled}),
\begin{eqnarray}
n_1(t, \rho)=\frac{dN_e}{d\zeta}\frac{1}{4\pi D_1 t}\exp\left(-\frac{\rho^2}{4 D_1 t}-\frac{t}{\tau_1} \right),\nonumber\\
n_2(t,\rho)=2\frac{1}{4\pi\tau_1} \frac{dN_e}{d\zeta}\int_0^t \frac{1}{D_1 t'+D_2(t-t')} \nonumber \\ \times \exp\left(-\frac{\rho^2}{4( D_1 t'+D_2(t-t'))}-\frac{t-t'}{\tau_{2}}-\frac{t'}{\tau_1}\right)dt'.
\label{integrand3E}
\end{eqnarray}
Notice, that $n_2(t,\rho)$ is doubled since it comprises the newly ejected electrons as well as the secondary electrons of the first generation that lost energy.
The results for these number densities are shown in Fig.~\ref{fig:ele}.
From this figure, it is clear that after a few fs after the ion's passage, all secondary electrons lose energy and the number density of secondary electrons is by and large is given by $n_2(\rho,t)$. In Fig.~\ref{fig.el2}, $n_2(\rho,t)$ is plotted at different times as a function of distance from the path. With time the distribution becomes a little broader, but the main effect is the exponential decrease with time.
As a result of these decrease, the so-called pre-solvated electrons are formed. A pre-solvated stage of electrons is a transition stage between low-energy ballistic electrons and a relatively stable compound of electrons with water molecules known as solvated electrons. This transition takes about 1~ps.



\begin{figure}
\begin{centering}
\resizebox{0.99\columnwidth}{!}
{\includegraphics{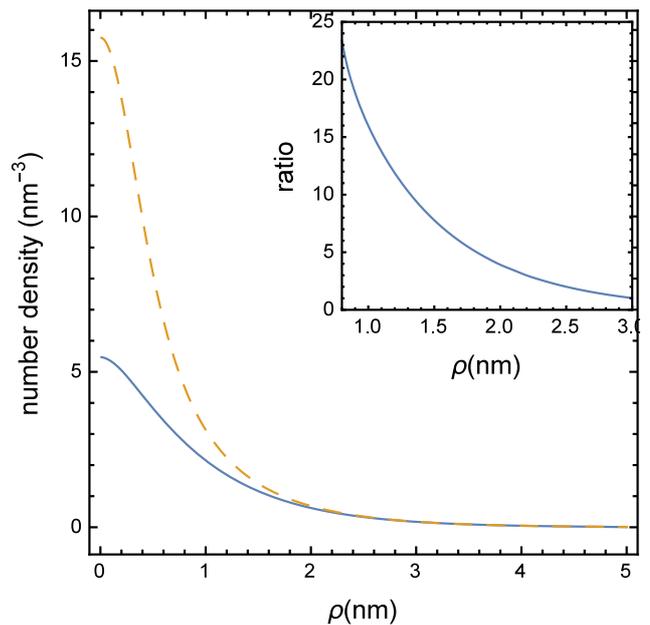}}
\caption{\label{fig:aq} (Colour online)
Number density of pre-solvated electrons (solid line) and initial distribution of hydroxyl radicals (dashed line) as a function of distance from the ion's path at 50~fs when the transport of ballistic electrons is over. The ratio of chemical annihilation of hydroxyl radicals to their diffusion as a function of the distance from the ion's path is shown in the inset.}
\end{centering}
\end{figure}

The formation of pre-solvated electrons can be found from the conservation of electrons corresponding to~(\ref{coupledE}),
\begin{eqnarray}
\frac{\partial n_{aq}({\bf r}, t)}{\partial t} = \frac{n_{2}({\bf r}, t)}{\tau_{2}},
\label{eaq}
\end{eqnarray}
where the subscript ``{\em aq}'' corresponds to pre-solvated electrons.
In~(\ref{eaq}), the diffusion and all chemical terms are dropped since the diffusion ($D_{aq}=4.5\times 10^{-6}$nm$^2$fs$^{-1}$~\cite{laverne89}) and chemical reactions are happening on a much longer scale. Equation~(\ref{eaq}) can be integrated over time to get the initial distribution of pre-solvated electrons.
During the time of formation of solvated electrons, pre-solvated electrons may interact with nearby biomolecules, but, since their diffusion coefficient is much smaller compared to those of ballistic electrons, their location does not change considerably compared to the position of last inelastic event. Therefore, the distribution obtained by the integration of (\ref{eaq}) is the initial radial distribution of solvated electrons.
This distribution is shown in Fig.~\ref{fig:aq}.
Later on they slowly diffuse and react with other solvated electrons and  hydroxyl radicals to form stable OH$^-$~\cite{laverne89}. Chemical reactions  dominate the diffusion and, as will be discussed below, given that diffusion is the only mechanism for the transport of solvated electrons, they are unlikely to leave the track.


It is also possible to assess the energy deposition density (dose), $\varepsilon(\rho,t)$, in the medium by ions and secondary electrons. In order to do this, we assume that the average energy, ${\bar w}$, is deposited to the medium with each ionization. Then, the rate of energy deposition is proportional to the rate of inelastic events,
\begin{eqnarray}
\frac{\partial\varepsilon(\rho,t)}{\partial t}={\bar w}\left(\frac{dN_e}{d\zeta}\delta^{(2)}(\rho)\delta(t)+\frac{n_{1}({\rho}, t)}{\tau_{1}}+\frac{n_{2}({\rho}, t)}{\tau_{2}}\right).
\label{en.depos}
\end{eqnarray}
The first energy deposition occurs right at the ion's path where the molecules are ionized by the ion; this corresponds to the first term on the right hand side of (\ref{en.depos}), where $\delta$'s are the corresponding $\delta$-functions. The second deposition (second term on the r.h.s.) is the ionization by secondary electrons at the end of their ionization mean free paths. Finally, the third deposition is due to remaining energy loss due to excitation of molecules by the electrons of second generation. After that, the electrons enter the pre-solvated stage.

\begin{figure}
\begin{centering}
\resizebox{0.99\columnwidth}{!}
{\includegraphics{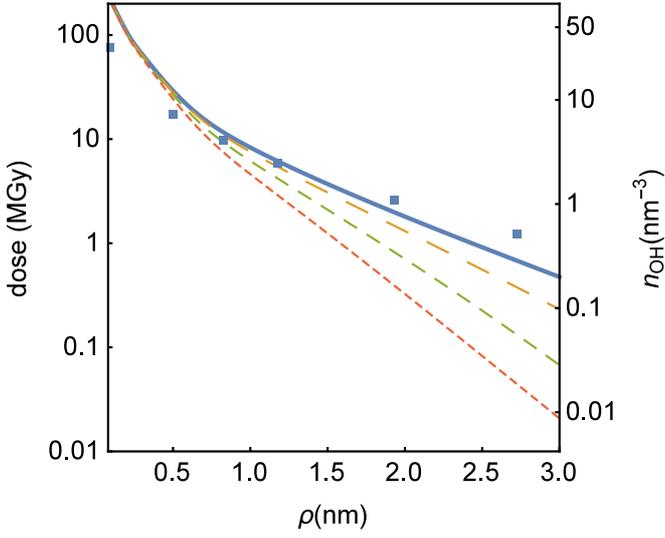}}
\caption{\label{fig:dose} (Colour online)
The radial dose (in MGy) distributions as functions of the distance from the ion's path at times 50, 20, 10, and 5~fs are shown with solid and dashed lines with a diminishing dash size, correspondingly. The solid line also represents the initial distribution of hydroxyl radicals. The corresponding labels are shown on the right side of the frame. The dots mark the radial dose data for 2-MeV/u~\cite{31} multiplied by the factor of four.}
\end{centering}
\end{figure}

The time integration of~(\ref{en.depos}) gives the dependence of the radial dose on time. The radial dose distributions at times 5, 10, 20, and 50~fs are shown in Fig.~\ref{fig:dose}. At small radii, the distribution is due to primary ionization and it does not change with time. At larger radii, the dose slowly increases because of energy loss by the second generation of electrons. The shown results are obtained with ${\bar w}=16.5$~eV, which corresponds to the normalization
\begin{eqnarray}
\int\int\frac{\partial\varepsilon(\rho,t)}{\partial t}dt 2\pi\rho d\rho=S_e,
\label{en.norm}
\end{eqnarray}
where $S_e=900$eV/nm is the linear energy transfer (LET) of a {\em single ion} at the Bragg peak for carbon ions~\cite{epjd}.
In Fig.~\ref{fig:dose}, the results of integration of (\ref{en.depos}) are compared with the radial dose of Ref.~\cite{31}. While the shapes of these distributions are alike, the
absolute values are different by the factor of about four. Part of this disagreement (factor of 1.9) can be explained by the fact that in~\cite{31}, the radial dose presented for 2-MeV/u, i.e., proximal of the Bragg peak for carbon ions. The remaining factor can be due to the energy straggling present in Ref.~\cite{31}, but absent for a single ion data of~(\ref{en.depos}). Also, while the solid line represents the radial dose at 50~fs, in Ref.~\cite{31} there is no information about the time. This is typical for Monte Carlo simulations, but since they are compared with experimental data the time corresponding to dots is likely to be on a ps scale.

The results shown in Fig.~\ref{fig:dose} may be used as a starting point for shock wave development~\cite{MSAColl,natnuke}, since the radial dose distribution evolved by the time $\leq 50$~fs determines the temperature distribution around the ion's path following from the process of energy relaxation, in which the energy stored in electronic excitations is transferred into vibrational excitations and then to translational degrees of freedom. The characteristic time for the decay of an electronic excitation in Na clusters is estimated to be about 0.4~ps~\cite{gerchikov2000} and it is likely to be several times longer for the liquid water, which is consistent with the analysis performed in Ref.~\cite{preheat}. The dependence shown in Fig.~\ref{fig:dose} is somewhat similar to that discussed in Ref.~\cite{preheat}, but the difference is that (\ref{en.depos}) includes diffusion of secondary electrons, but does not include thermal conductivity, while Ref.~\cite{preheat} does the opposite.

Finally, the radial dose has been studied in Ref.~\cite{epjdisacc2011} using a random walk approximation. In that work, we obtained a reasonably good comparison of shapes with Ref.~\cite{31} for intermediate distances. Only one generation of secondary electrons was used, but the parameters, such as mean free path and the relaxation time were chosen close to those for the second generation of this work. Therefore, the results obtained in Ref.~\cite{epjdisacc2011} are meaningful, however, they do not include term ${\bar w}\frac{dN_e}{d\zeta}\delta^{(2)}(\rho)$ that comes from primary ionization events.

\section{Transport and production of hydroxyl radicals}

Hydroxyl among other radicals plays the most important role in the DNA damage~\cite{hyd2,Nikjoo97}. Therefore, we are considering the production and transport of these radicals in more detail. In the liquid water environment irradiated with ions such as carbon, the hydroxyl radicals are primarily produced as a consequence of ionization of water molecules,
\begin{eqnarray}
{\rm
H_2O^+ + H_2O \rightarrow H_3O^+ + OH^\cdot,
}
\label{rea:ionOH}
\end{eqnarray}
and by excited water molecules,
\begin{eqnarray}
{\rm
H_2O^\ast\rightarrow H^\cdot + OH^\cdot }.
\label{rea:excOH}
\end{eqnarray}
Both of these processes can be initiated by either ions or secondary electrons that ionize and excite water molecules. The time scales play an important role in further analysis. The ionization is happening in $10^{-17}$s and secondary electrons diffuse by several nm within several fs. However, ${\rm OH}^\cdot$ is produced through~(\ref{rea:ionOH}) and~(\ref{rea:excOH}) only on a ps scale, i.e., after the electron transport is over and pre-solvated electrons are formed. Neither ${\rm H_2O}^+$, nor ${\rm H_2O}^\ast$ are remaining at rest during this time; they are diffusing away from the path. Since their diffusion coefficients are of the order of $10^{-6}$nm$^2$fs$^{-1}$, they diffuse by only about $\sqrt{D \tau_{form}}\approx \sqrt{10^{-6} 10^3}=3\times 10^{-2}$nm away from places where the inelastic event (ionization or excitation of water molecule) has taken place. This distance is too small, therefore, we can consider the loci of inelastic events to be the initial loci of hydroxyl radicals.

The distribution of in elastic events was analysed in Sec.~\ref{sec.tertiaryelectrons} in Eq.~(\ref{en.depos}), where the rate of energy deposition was obtained. In the first approximation, we can assume that each inelastic event produces a hydroxyl radical, then the equation, similar to~(\ref{en.depos}) for production of radicals is as follows,
\begin{eqnarray}
\frac{\partial n_{\rm OH}(\rho,t)}{\partial t}=\frac{dN_e}{d\zeta}\delta^{(2)}(\rho)\delta(t)+\frac{n_{1}({\rho}, t)}{\tau_{1}}+\frac{n_{2}({\rho}, t)}{\tau_{2}}.
\label{oh.dist}
\end{eqnarray}
Similarly to~(\ref{en.depos}), it can be integrated over time to obtain the initial number density of hydroxyl radicals, $n_{\rm OH}(\rho)$. It is shown with a solid line in Fig.~\ref{fig:dose} with labels on the right side of the graph. The results of the integration of second and third terms are shown with a dashed line in Fig.~\ref{fig:aq} to compare the distributions of hydroxyl with solvated electrons at larger distances.

Thus, with the above assumptions, each ionization on the ion's path produces one 45-eV electron, which after another ionization loses energy and produces two 15-eV electrons, which then become two solvated electrons. These two ionization events together with final energy loss by two 15-eV electrons produce four hydroxyl radicals. However, the initial radial distributions of solvated electrons and hydroxyl are different, the hydroxyl strongly dominates near the path and they are the same at $\rho>1.5$nm. Now, let us analyse the further evolution of hydroxyl.


As can be seen from Fig.~\ref{fig.el2}, the ballistic electrons are gone after about 50~fs, a time much shorter than that of formation of hydroxyl. Thus, by the time hydroxyl radicals are formed, there are no more active sources since both ionization and excitation processes are over. All reactions that eradicate hydroxyl are second order reactions with respect to perturbations in water; these are
\begin{eqnarray}
{\rm OH^\cdot + H^\cdot \rightarrow  H_2O,} \nonumber\\
{\rm OH^\cdot} + e^-_{aq} \rightarrow {\rm OH^-}, \nonumber\\
{\rm OH^\cdot + OH^\cdot \rightarrow H_2O_2,
}
\label{rea:OHrem}
\end{eqnarray}
and others~\cite{laverne89}. This means that all equations such as~(\ref{raddiff}) and~(\ref{coupled}) acquire additional terms on the right hand side that account for chemical reactions, but lose terms like ${n_{\rm OH}({\bf r}, t)}/{\tau_{\rm OH}}$, since hydroxyl does not react with water molecules and is stable by itself. For example, Eq.~(\ref{raddiff}) becomes
\begin{eqnarray}
\frac{\partial n_{\rm OH}({\bf r}, t)}{\partial t} = D_{\rm OH} \nabla^2 n_{\rm OH}({\bf r}, t)-\sum k_{{\rm OH}i}n_{\rm OH} n_i, \label{raddiffchem}
\end{eqnarray}
where $k_{{\rm OH}i}$ is the rate constant corresponding to the reaction of species ``$i$'' with hydroxyl. This is a nonlinear equation, and we start with comparing all terms on the right hand side of (\ref{raddiffchem}).

First, let us consider the hydroxyl radicals that are formed due to ionizations or excitations by ions.
Since the diffusion coefficient for hydroxyl is so small ($2.8\times 10^{-6}$~nm$^2$fs$^{-1}$) compared to electrons discussed above, the first term in Eq.~(\ref{raddiffchem}) (for carbon ions in the vicinity of Bragg peak) is of the order of 10 if we consider the estimated diffusion radius to be 0.03-nm. The second term is equal to $k_{{\rm OH,OH}}n_{\rm OH}^2$, since hydroxyl by itself is the dominant reagent in this region. With $k_{{\rm OH,OH}}=1.0\times10^{-5}$nm$^{3}$fs$^{-1}$~\cite{laverne89,Nikjoo97,hyd2}, this term exceeds $200$nm$^{-3}$fs$^{-1}$, since the concentration near the path is so high. This concentration decreases by the factor of 100 within 2~ps as most of the hydroxyl becomes involved in the formation of peroxide. Therefore, the hydroxyl, formed by direct ion's action by and large does not exit the a sub-nm region around the ion's path.

The situation with hydroxyl formed by secondary electrons is somewhat different, since it is formed with a much smaller concentration and the diffusion term may be comparable with that of peroxide formation. The equation of interest can be written as,
\begin{eqnarray}
\frac{\partial n_{\rm OH}(\rho, t)}{\partial t} = D_{\rm OH} \frac{1}{\rho}\frac{\partial}{\partial \rho} \left(\rho \frac {\partial n_{\rm OH}(\rho, t)}{\partial \rho}\right) \nonumber \\- k_{{\rm OH},e_{aq}}n_{\rm OH}(\rho, t)n_{aq}(\rho, t)- k_{{\rm OH,OH}}n_{\rm OH}^2(\rho, t). \label{secOH}
\end{eqnarray}
This equation includes two relevant chemical reactions from~(\ref{rea:OHrem}). In the beginning, at $t\approx 1$ps, the number density of hydroxyl, $n_{\rm OH}(\rho, t)$, exceeds that of solvated electrons, $n_{aq}(\rho, t)$, as can be seen in Fig.~\ref{fig:aq}. The reaction constant $k_{{\rm OH},e_{aq}}$ is larger than $k_{{\rm OH,OH}}$ by the factor of 4.2~\cite{laverne89,Nikjoo97,hyd2}. Therefore, the reaction terms in~(\ref{secOH}) dominate the diffusion term in the whole domain of $\rho$. The most conservative estimate of the ratio of reaction to diffusion terms in~(\ref{secOH}) is shown in the inset of Fig.~\ref{fig:aq}. It is apparent that terms become comparable only at $\rho \ge 2.5$~nm. Only a small (less than 10\%) fraction of hydroxyl diffuses to these distances.

Therefore, we can summarise that in the first approximation, which can be improved by introduction of probabilities of formation of hydroxyl radicals, the initial distribution of OH$^\cdot$ is shown in Figs.~\ref{fig:aq}-\ref{fig:dose}. These radicals would react with biomolecules such as DNA if the latter appear to be in the track. However, if the collective transport with a shock wave does not take place, the radicals react with each other and with solvated electrons and by and large do not leave the track. Other species, such as ${\rm H_2O_2}$ and ${\rm OH^-}$ may propagate on larger distances and may be detected outside tracks.



\section{Conclusion}

The random walk approximation, applied to two generations of secondary electrons showed that most of them do not spread beyond the 2-nm cylinder around the ion's path. Pre-solvated electrons, important agents of strand breaks in DNA molecules, are also formed within this small region. Their small diffusion coefficients do not allow them to be transferred far enough since they readily react with hydroxyl radicals abundant in the same region. The latter also react with themselves and their high number density makes this (nonlinear) reaction much more important than their (linear) diffusion. Thus, we can conclude that the diffusion mechanism does not allow most of the reactive species to leave the few-nm track radius. In this work, carbon ions near the Bragg peak were used as an example. The effect of containment of reactive species in the track will be stronger for heavier ions and weaker for lighter ones, since the concentration of reactive species is much smaller.

Alternatively, another transport mechanism, i.e., collective transport in a shock wave should be studied both theoretically and experimentally in order to understand whether the reactive species actually propagate to larger distances. This question is very important for the assessment of radiation damage with ions~\cite{MSAColl}. The radial dose distribution obtained in Fig.~\ref{fig:dose} gives the initial conditions for the development of cylindrical shock waves.

\section*{Acknowledgements}
We are grateful to A.~Adhikary and M.~D. Sevilla for their advice and fruitful discussions,
and the support of COST Action MP1002 ``Nano-scale insights in ion beam cancer therapy.''

\bibliographystyle{epj}

\end{document}